\documentclass[10pt]{article}
\usepackage{graphicx}
\usepackage{lineno}

 \pdfoutput=1 

\def\Title#1{\begin{center} {\Large #1 } \end{center}}
\def\Author#1{\begin{center}{ \sc #1} \end{center}}
\def\Address#1{\begin{center}{ \it #1} \end{center}}

\newcommand\pubblock{\rightline{\begin{tabular}{l} Proceedings of the Fifth Annual LHCP\\ \pubnumber\\
         \pubdate  \end{tabular}}}

\newenvironment{Abstract}{\begin{quotation} \begin{center} 
             \large ABSTRACT \end{center}\bigskip 
      \begin{center}\begin{large}}{\end{large}\end{center} \end{quotation}}

\newenvironment{Presented}{\begin{quotation} \begin{center} 
             PRESENTED AT\end{center}\bigskip 
      \begin{center}\begin{large}}{\end{large}\end{center} \end{quotation}}

\def\beq{\begin{equation}}
\def\eeq#1{\label{#1}\end{equation}}
\def\eeqn{\end{equation}}

\def\beqa{\begin{eqnarray}}
\def\eeqa#1{\label{#1}\end{eqnarray}}
\def\eeqan{\end{eqnarray}}

\let\bar=\overbar

\def\Dslash{\not{\hbox{\kern-4pt $D$}}}
\def\dslash{\not{\hbox{\kern-2pt $\del$}}}

\def\msb{{\bar{\ssstyle M \kern -1pt S}}}

\usepackage{color}

\newcommand\pubnumber{ CMS CR-2017/225 }

\newcommand\pubdate{\today}

\def\affiliation{
On behalf of the ATLAS and CMS Collaborations, \\
Instituto de F\'isica de Cantabria \\
CSIC - Universidad de Cantabria, Santander, 39005, Spain}

\begin{document}

\large
\begin{titlepage}
\pubblock

\vfill
\Title{  b TAGGING IN ATLAS AND CMS  }
\vfill

\Author{ LUCA SCODELLARO  }
\Address{\affiliation}
\vfill
\begin{Abstract}

Many physics signals presently studied at the high energy collision experiments lead to final states with jets originating from heavy flavor quarks. This report reviews the algorithms for heavy flavor jets identification developed by the ATLAS and CMS Collaborations in view of the Run2 data taking period at the Large Hadron Collider. The improvements of the algorithms used in 2015 and 2016 data analyses with respect to previous data taking periods are discussed, as well as the ongoing developments in view of the next years of data taking. The measurements of the performance of the algorithms on data as well as the dedicated techniques for the identification of heavy flavor jets in events with boosted topologies are also presented. Finally, the effectiveness of heavy flavor jet identification in the complex environment expected during the high luminosity LHC phase is discussed.

\end{Abstract}
\vfill

\begin{Presented}
The Fifth Annual Conference\\
 on Large Hadron Collider Physics \\
Shanghai Jiao Tong University, Shanghai, China\\ 
May 15-20, 2017
\end{Presented}
\vfill
\end{titlepage}
\def\thefootnote{\fnsymbol{footnote}}
\setcounter{footnote}{0}
%

\normalsize 



\section{Introduction}

The identification of jets originating from the hadronization of heavy flavor quarks is a key ingredient of many high precision measurements and searches for new physics beyond the standard model (SM) at the Large Hadron Collider (LHC). 
Both the ATLAS \cite{ATLASDet} and CMS \cite{CMSDet} Collaborations have developed efficient identification algorithms during the LHC Run1 data taking period \cite{Run1ATLAS,Run1CMS}. This report presents an overview of the latest developments of the heavy flavor identification algorithms (Section \ref{sec:algorithms}), and of the measurements of their performance on data (Section \ref{sec:performance}). Focus is also given to the extension of the b tagging techniques to events with boosted topologies (Section \ref{sec:boosted}) and to the preparation of the algorithms to the challenges of the future runs of data taking (Section \ref{sec:upgrades}).

\section{b Jet Identification Algorithms}\label{sec:algorithms}

The identification of the jets originating from the hadronization of heavy flavor quarks is made possible by the distinctive properties of the heavy hadrons produced in the process. For instance, their large lifetime allows them to travel a measurable distance from the primary interaction point before to decay, giving rise to displaced tracks which can form secondary vertices. Their high mass also leads to decay products with a larger transverse momentum relative to the jet axis with respect to the ones typically found in jets from light partons. Finally, heavy hadrons have a sizable branching ratio for semileptonic decays, hence the presence of soft leptons in the produced jets provides another tool for heavy jet identification.

The ATLAS and CMS Collaborations developed several algorithms to identify (tag) the jets from b quark hadronization based on the properties detailed above. The general strategy is to start with simple algorithms that exploits a particular property of b jets and progressively add more information to build more sophisticated algorithms. The most performant algorithms presently in use in physics analyses at ATLAS and CMS are based on multivariate (MVA) combinations of the available information. The output of these algorithms consists in a discriminant value for each jet. Operating points are then defined as thresholds on the discriminant, designed to provide a determined efficiency for identifying b jets (in ATLAS) or to reduce the probability of mis-tagging a light jet to a prefixed level (in CMS).

\begin{figure}[htb]
\centering
\includegraphics[height=2in]{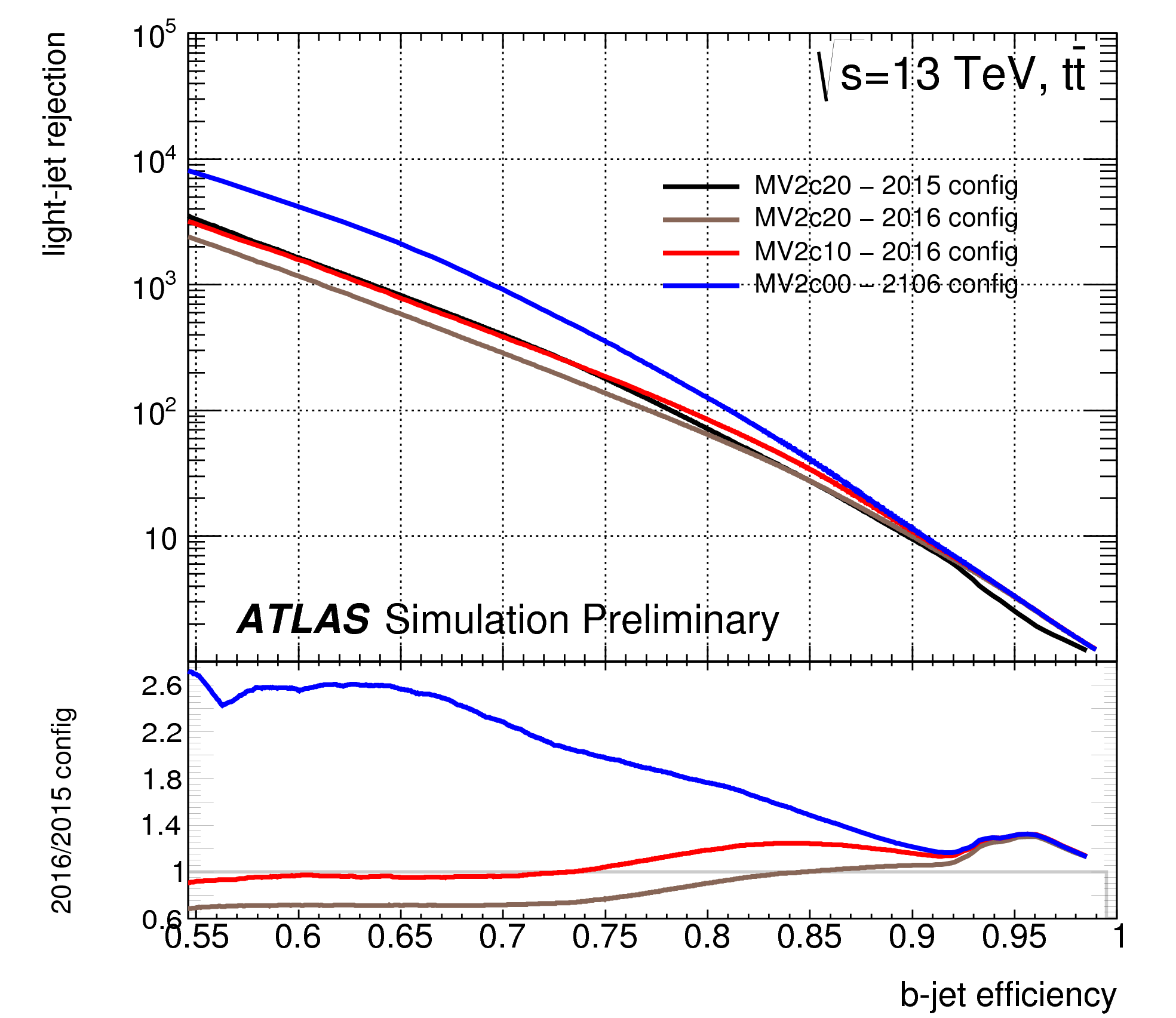}
\includegraphics[height=2in]{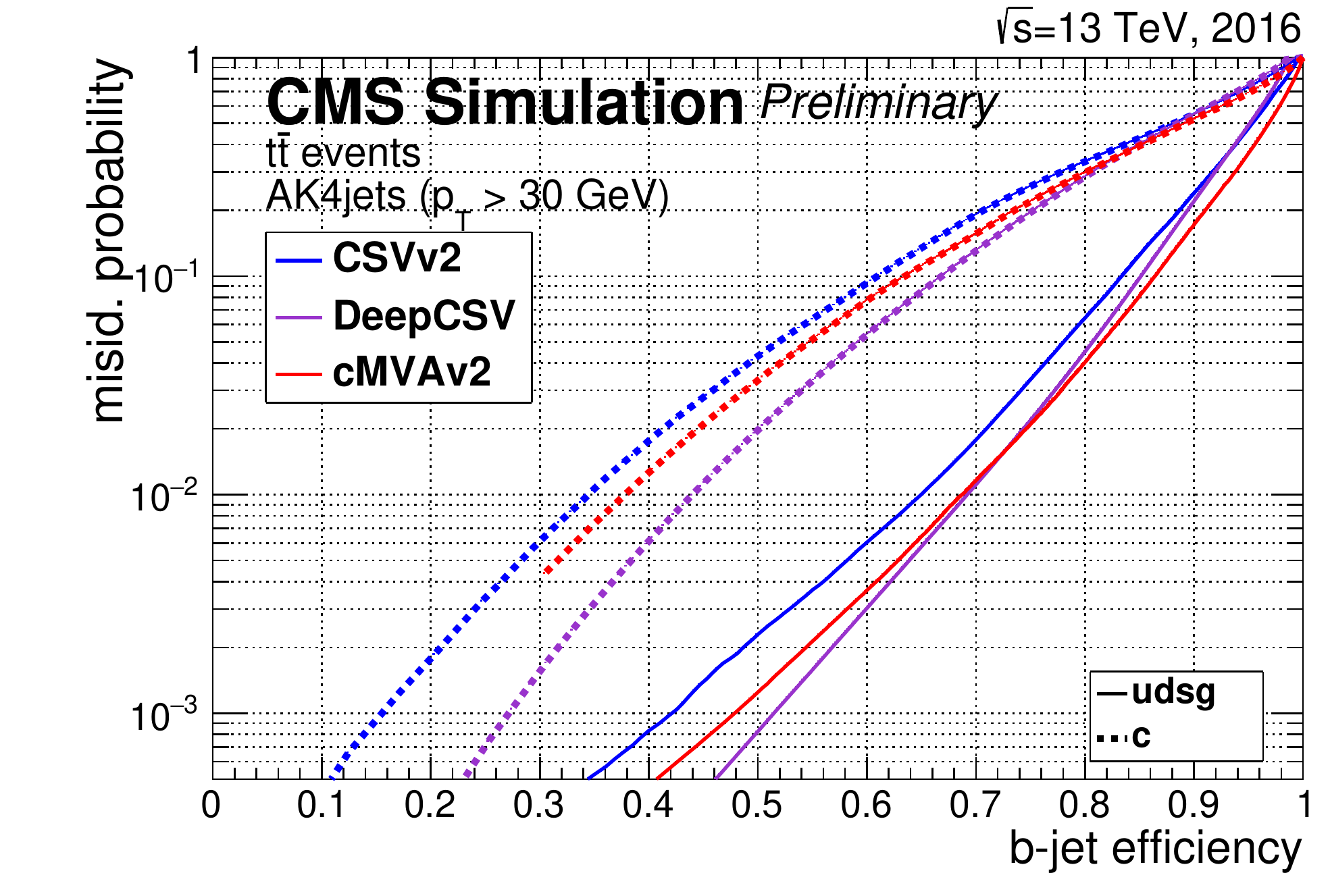}
\caption{Performance of the main b jet identification algorithms used for physics analyses of 2015 and 2016 data at ATLAS \cite{ATLAS-ALG2} (left) and CMS \cite{CMS-ALG2} (right) experiments.}
\label{fig:algorithms}
\end{figure}

Figure \ref{fig:algorithms} shows the performances of the main b jet identification algorithms used for physics analyses of 2015 and 2016 data, represented for ATLAS (CMS) as curves of the light or c jet rejection factor (mis-identification probability) vs the b tagging efficiency. The Run2 algorithms in ATLAS benefited of improvements in the track detector and reconstruction algorithms \cite{ATLAS-ALG1} as well as of optimized criteria in the selection of input tracks and secondary vertices and improvements on the MVA training procedure \cite{ATLAS-ALG2}. The main algorithms in use at CMS are CSVv2, a neural network based evolution of the Run1 CSV algorithm which exploits a larger number of discriminating variables including a more inclusive secondary vertex finder, and cMVAv2, which combines the output from CSVv2 with the ones from other algorithms exploiting different b jets properties like the presence of soft leptons \cite{CMS-ALG1,CMS-ALG2}. 

\begin{figure}[htb]
\centering
\includegraphics[height=2in]{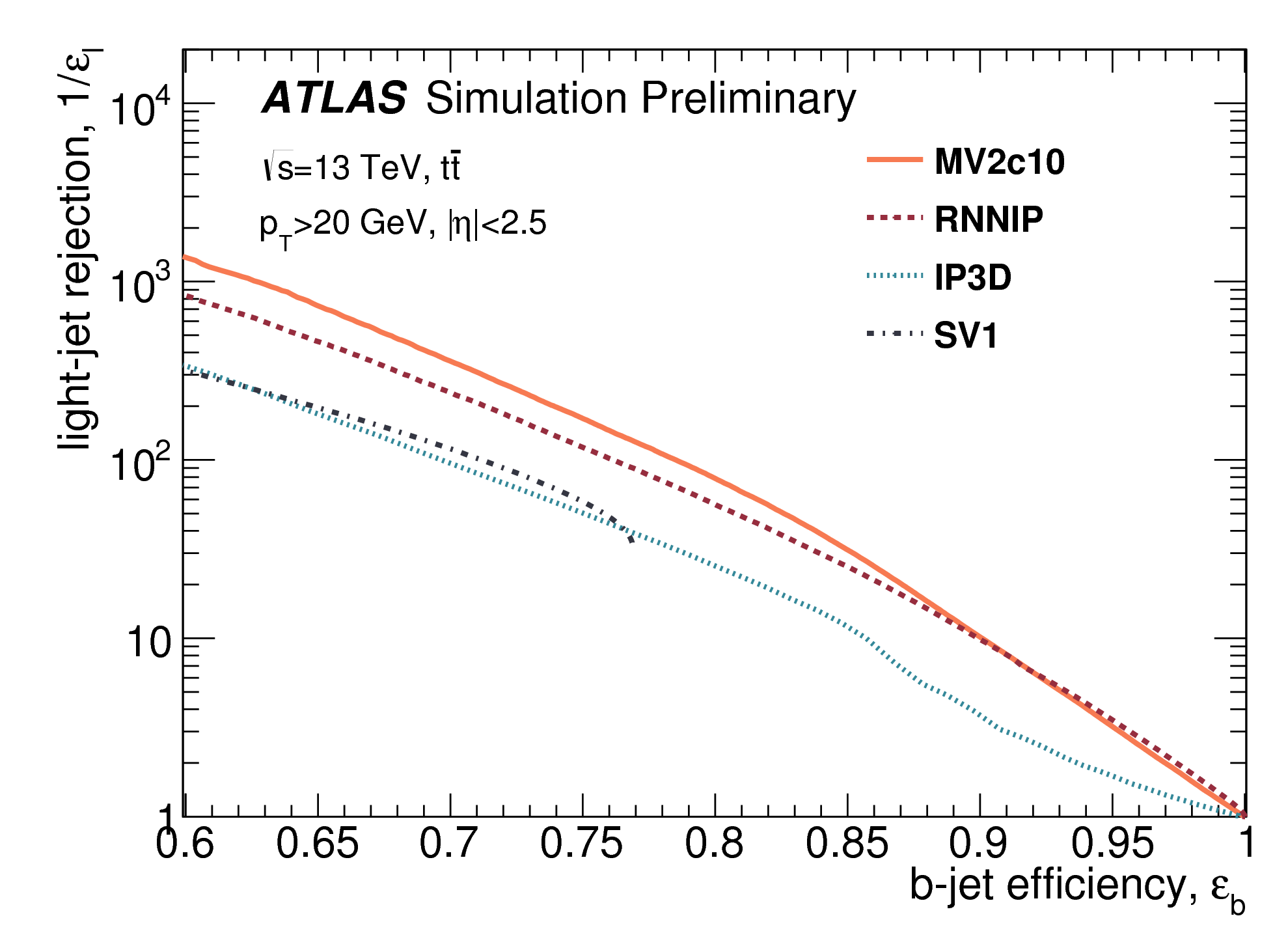}
\includegraphics[height=2in]{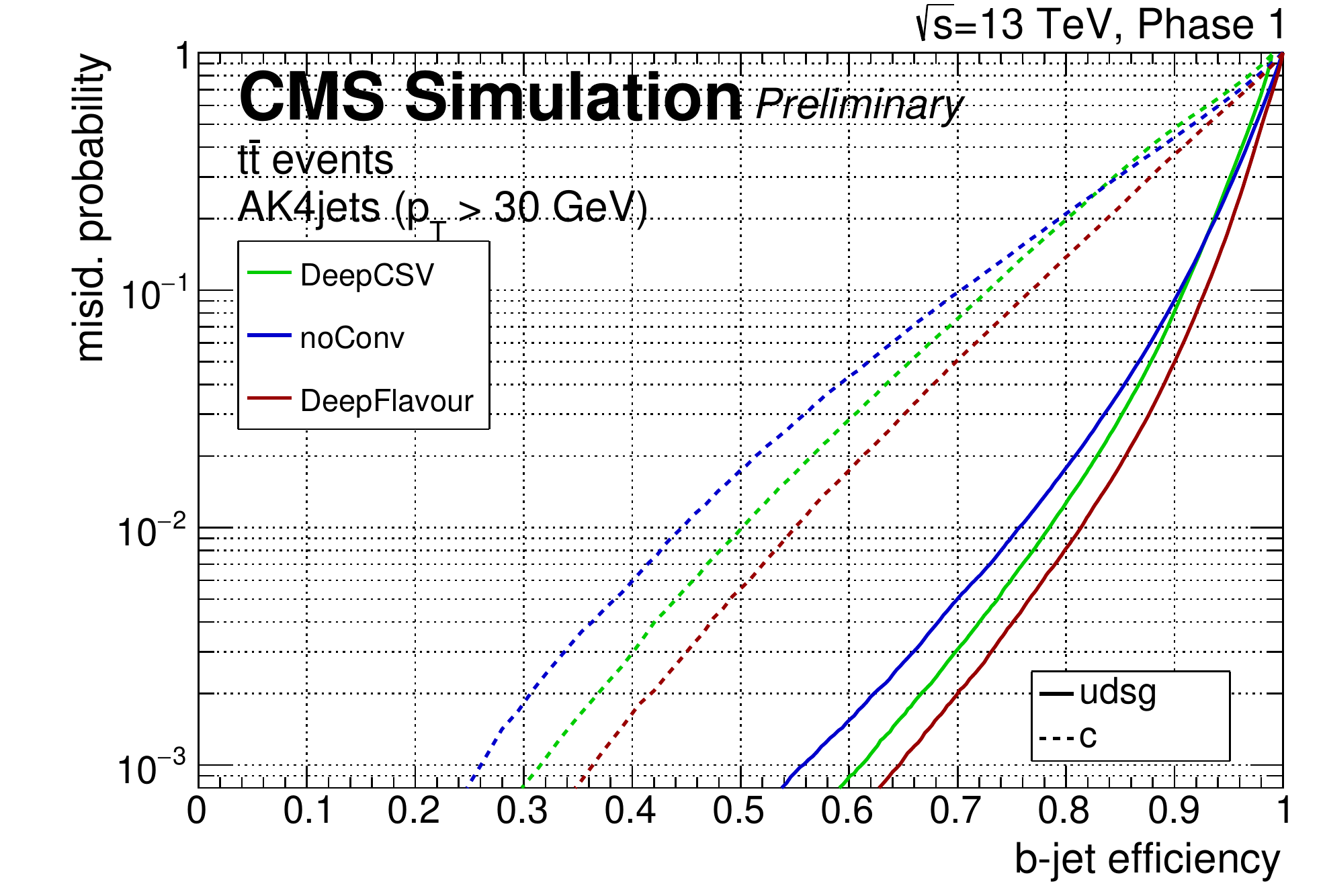}
\caption{Performance on new b jet identification algorithms based on recurrent neural networks at ATLAS \cite{ATLAS-ALG3} (left) and on deep neural networks at CMS \cite{CMS-ALG3} (right). CMS results are based on the new pixel detector installed in 2017.}
\label{fig:developments}
\end{figure}

The use of more sophisticated neural network classes allows to better exploit the information available to identify b jets for instance by combining a large number of input variables or making use of more low-level information. It also allows for multi-classification, providing an output probability for  each jet flavor hypothesis. The ATLAS Collaboration is developing new algorithms based on recurrent neural networks \cite{RNN}, whose directed cycles allow to process an arbitrary sequence of inputs. A RNNIP tagger \cite{ATLAS-ALG3} using just a sequence of track-by-track variables as input has already been found to outperform a standard algorithm based on track impact parameter information (see Figure \ref{fig:developments}, left). The CMS Collaboration exploited deep neural networks \cite{DNN} to build a DeepCSV algorithm \cite{CMS-ALG2} able to improve the performance of CSVv2 by using the same input variables but a larger set of tracks (see Figure \ref{fig:algorithms}, right). A new DeepFlavour tagger is further being developed to combine all the information on particles and secondary vertices in the jets \cite{CMS-ALG3}. Preliminary results show that a $4\%$ increase of efficiency with respect to DeepCSV for a mis-identification probability of $0.1\%$ is at reach, as shown in Figure \ref{fig:developments} (right). 



\begin{figure}[htb]
\centering
\includegraphics[height=2in]{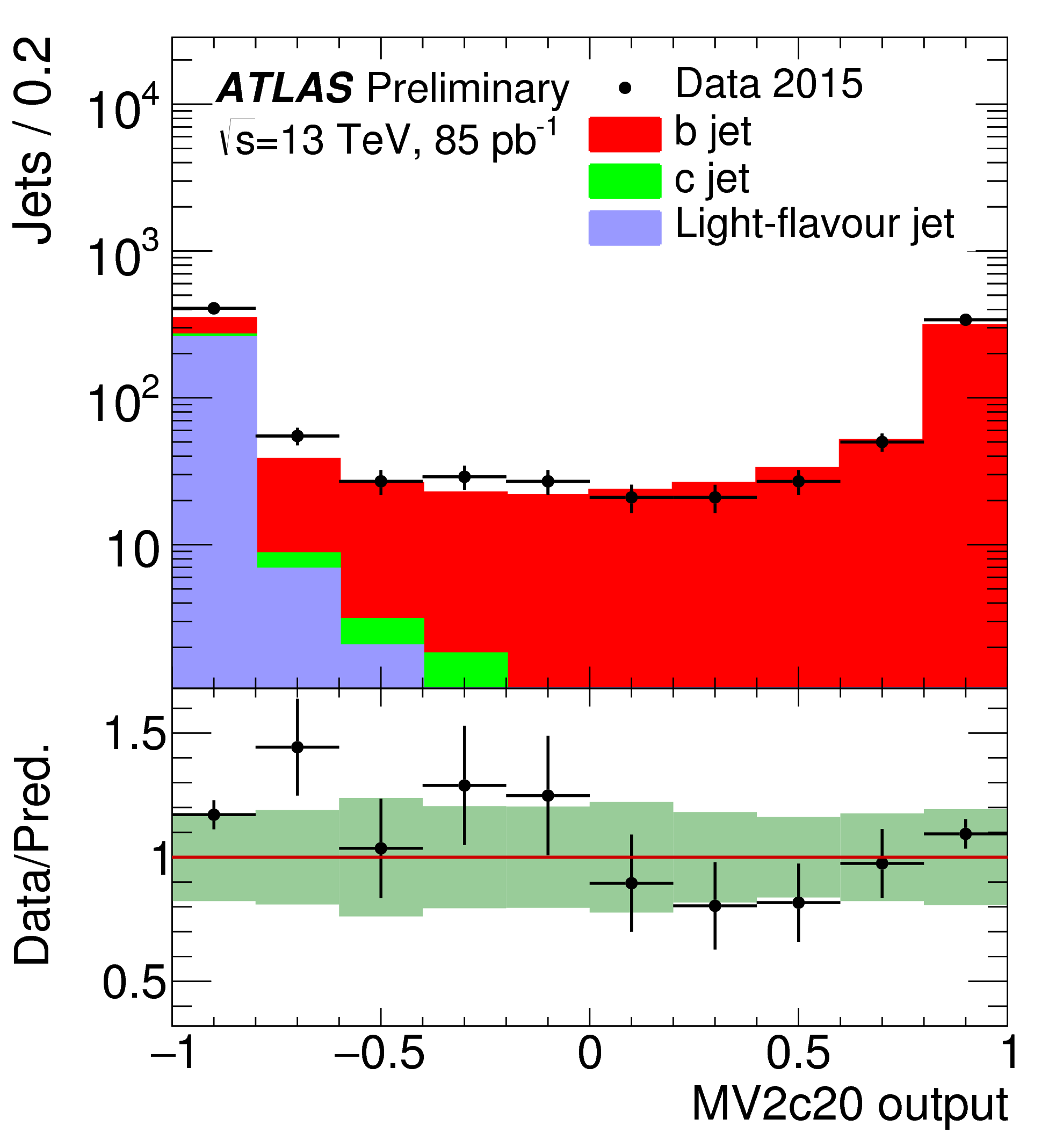}
\includegraphics[height=2in]{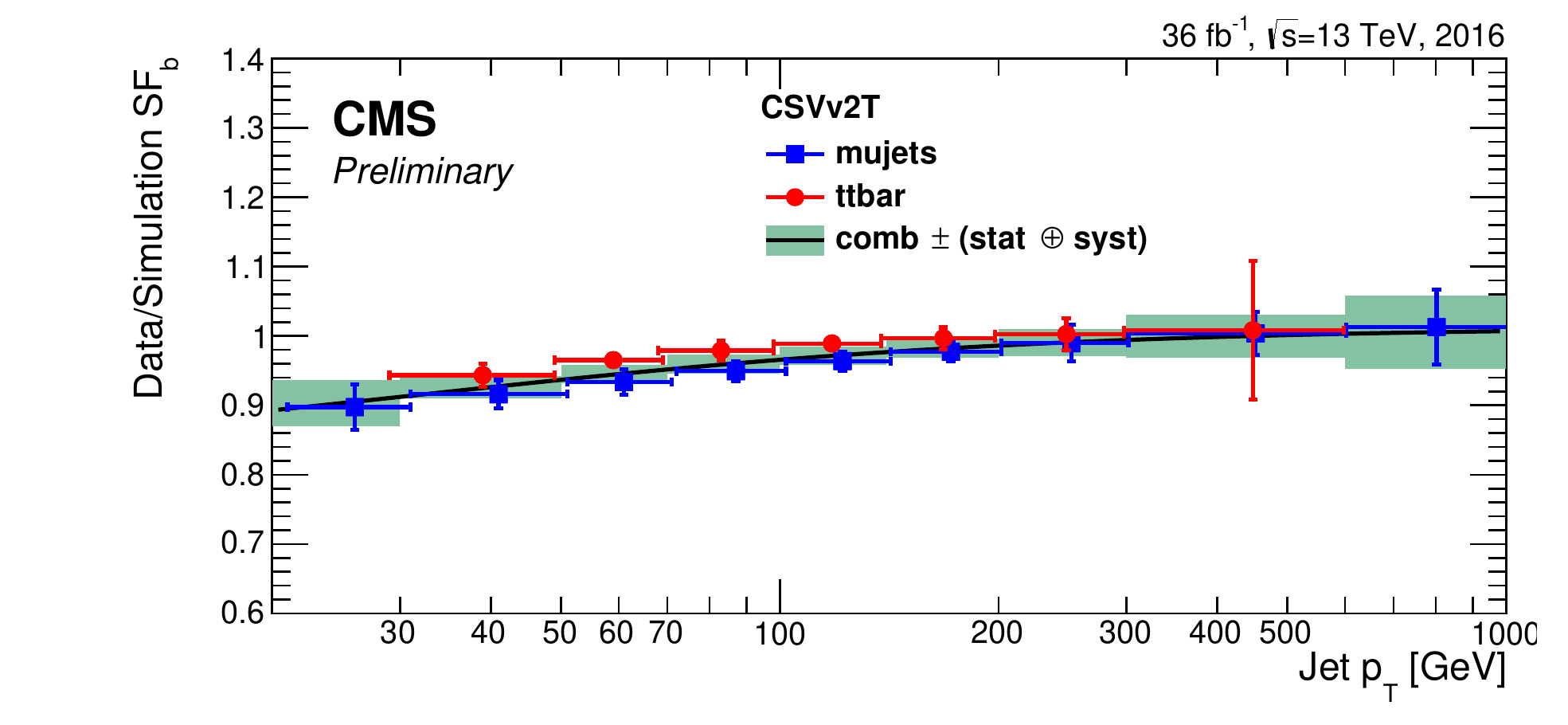}
\caption{Left: observed and expected distributions of the MV2c20 algorithm's output in top pair production events collected with the ATLAS detector \cite{ATLAS-COMM1}. Right: b tagging efficiency scale factors as a function of the jet transverse momentum measured by CMS in muon-enriched and $t\bar{t}$ events, and their combination \cite{CMS-PERF1}.}
\label{fig:performance}
\end{figure}

\section{Measurement of the Performance of the b tagging Algorithms}\label{sec:performance}

Monte Carlo simulations are not able to model very well the performance of the b tagging algorithms in data. For this reason, several samples with different topologies and heavy flavor content are used to measure the performance on data: inclusive jet samples from QCD processes allow to study the behavior of the algorithms in events dominated by light flavor jets; selecting jets with an embedded soft muon provides instead a sample of data enriched in heavy flavor jets; finally top pair production $t\bar{t}$ events  constitute a dataset very pure in b jets with low gluon splitting contribution \cite{CMS-ALG2,CMS-PERF1,ATLAS-COMM1}. Figure \ref{fig:performance} (left) shows the distribution of the MV2c20 algorithm´s output as observed in data and expected from simulations in $t\bar{t}$ events collected by the ATLAS experiment. 

Physics analyses will have to correct the expected behavior of b tag correlated observables as predicted by the simulations by the performance observed in data. For this purpose, data-to-MC scale factors are computed both for b tagging efficiency and mis-identification probability corresponding to each of the defined operating points of the algorithm. Mis-identification probability scale factors are measured in the inclusive jet samples, while b tagging efficiency scale factors are derived from muon-enriched jet samples and $t\bar{t}$ events \cite{CMS-PERF1,ATLAS-PERF1,ATLAS-PERF2}, and possibly combined, as shown in Figure \ref{fig:performance} (right).

\begin{figure}[htb]
\centering
\includegraphics[height=2in]{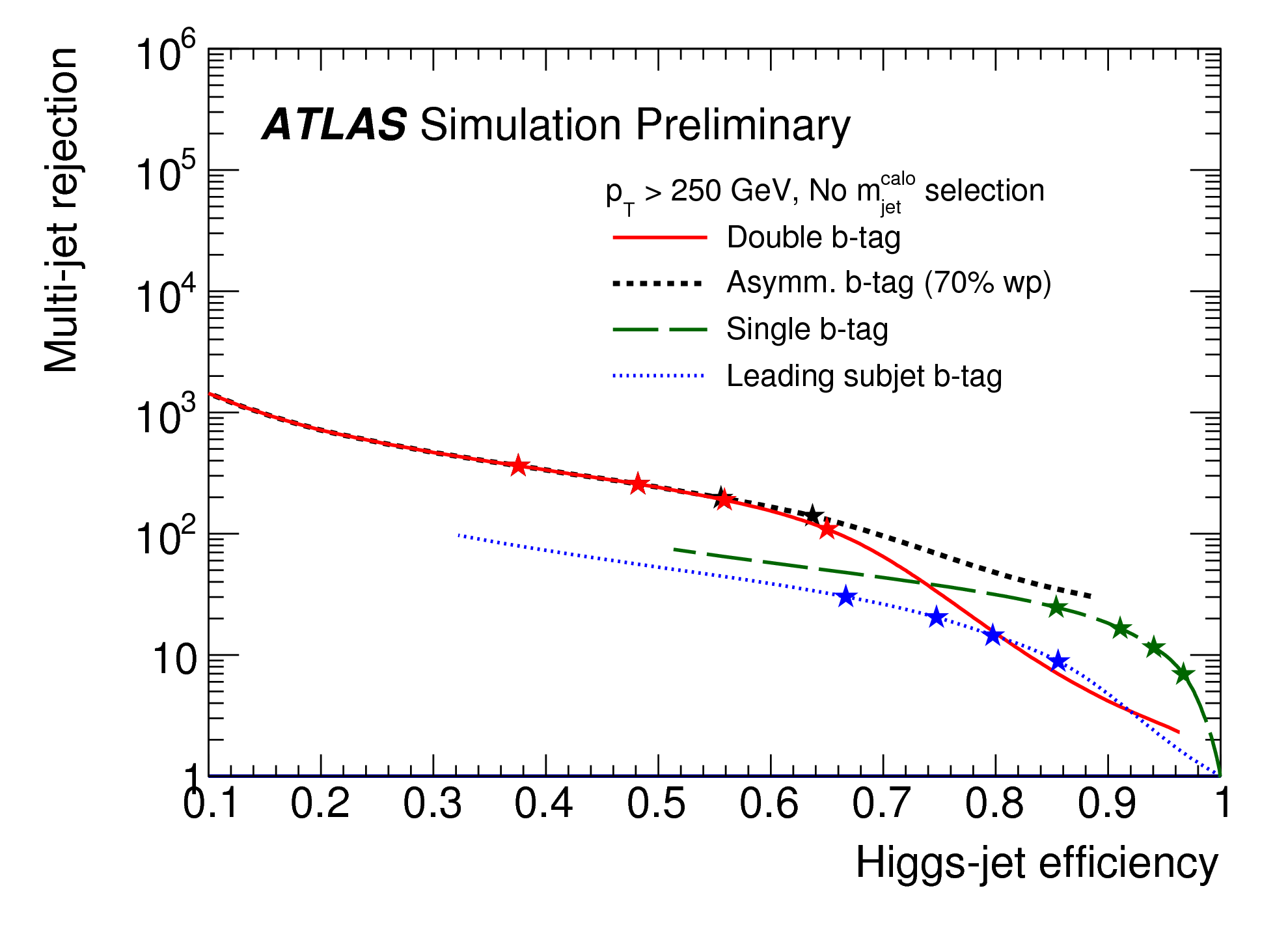}
\includegraphics[height=2in]{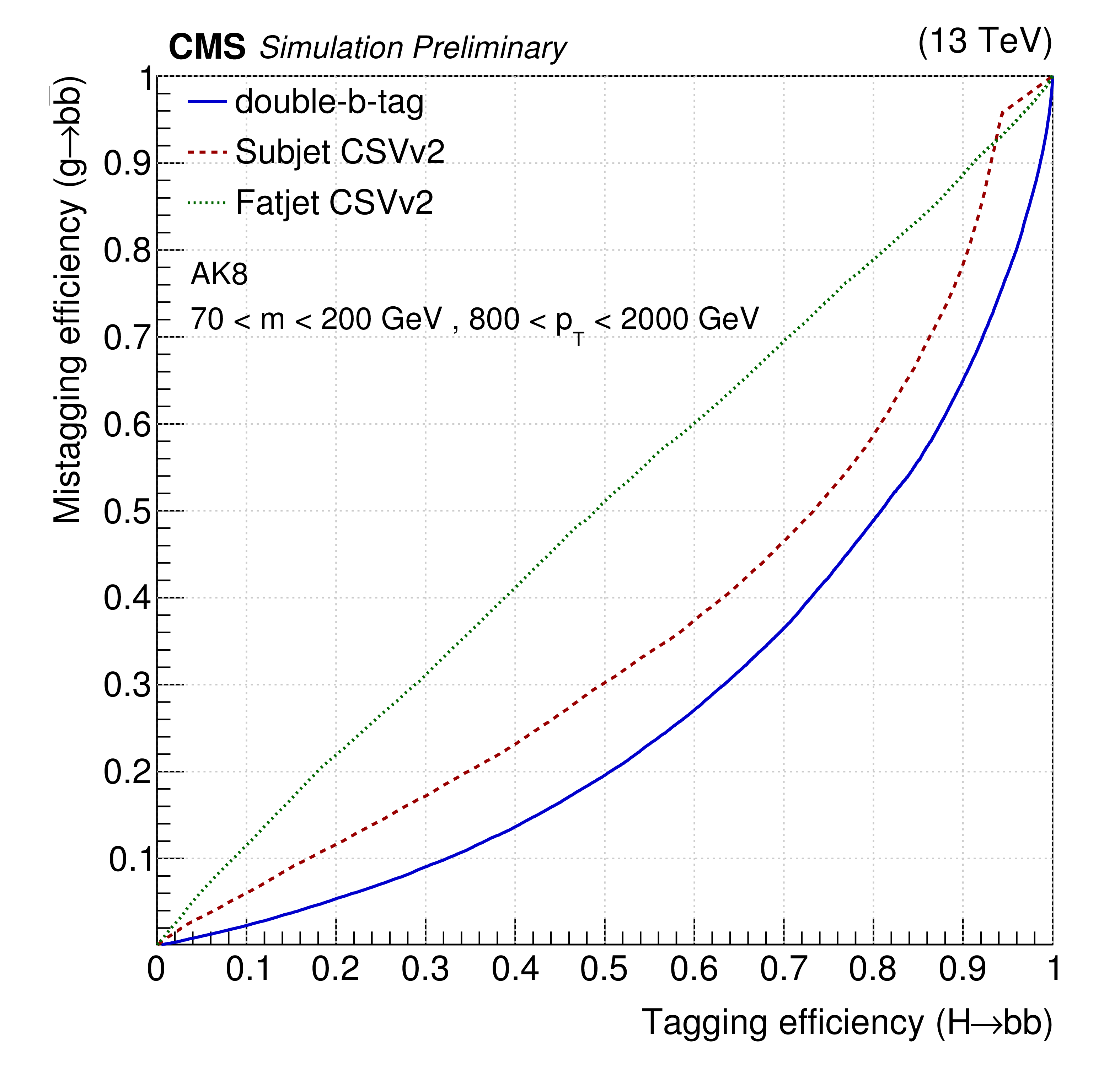}
\caption{Left: comparison of the performance for discriminating large jets from $H\rightarrow b\bar{b}$ decays against multijet background when different b tagging strategies are applied on the matched track jets \cite{ATLAS-BOOST1} (ATLAS). Right: comparison of the performance of the new double-b tagger and of fatjet and subjet tagging for identifying $H\rightarrow b\bar{b}$ decays against $g\rightarrow b\bar{b}$ decays \cite{CMS-BOOST1} (CMS).}
\label{fig:boosted}
\end{figure}

\section{b tagging in Events with Boosted Topologies}\label{sec:boosted}

In high energy collisions, particles decaying into b quarks can be produced with large momentum (boosted topology). The decay products of the B hadrons in the b jets can consequently overlap with the particles from other nearby jets. This situation arises for instance in the decays of heavy particles predicted by many models of physics beyond the SM. 

A dedicated boosted Higgs boson tagger has been developed by the ATLAS Collaboration for Run2 \cite{ATLAS-BOOST1}. Calorimetric jets with large cone size R=1 are used to reconstruct the boosted particles.
Ghost-association is then used to match the calorimetric jet with R=0.2 track jets, describing its decay products.
Finally, standard b tagging algorithms are applied on the track jets. Figure \ref{fig:boosted} compares various b tagging options to tag large calorimetric jets from $H\rightarrow b\bar{b}$ decays: in general, requiring two matched track jets satisfying (asymmetric) b tagging requirements is the most performant choice, with single track jet b tagging becoming competitive at high efficiencies. 
Additional requirements on the large calorimetric jet mass and substructure variables can be applied to enhance the performances.
The boosted Higgs boson tagger's performance is validated in a sample of candidates $g\rightarrow b\bar{b}$ large jets, where a good agreement of the observed and expected rates of double-b tagged large jets is found.

At the CMS experiment, candidate boosted particles are first reconstructed by particle flow AK8 jets (fatjets). Their decay products are then identified by resolving the jet substructure through soft drop declustering. Finally, b tagging algorithms can be applied on all the tracks in the fatjet or separately on the its subjets. While this last approach is still baseline in CMS searches involving boosted top quarks, a new dedicated algorithm to tag boosted decays to $b\bar{b}$ pairs has been developed in Run2 \cite{CMS-BOOST1}. The basic idea behind this new algorithm is to build observables from the n-subjettines axes to exploit the correlations between the two b quarks' flight directions. The double-b tagger has been found to outperform fatjet and subjet b tagging for $H\rightarrow b\bar{b}$ identification against multijets and $g\rightarrow b\bar{b}$ backgrounds (see Figure \ref{fig:boosted}, right). The performance of the new algorithm has been measured in data by applying standard technique to measure the efficiency scale factors to AK8 jets with two soft muon-tagged subjets.

\begin{figure}[htb]
\centering
\includegraphics[height=2in]{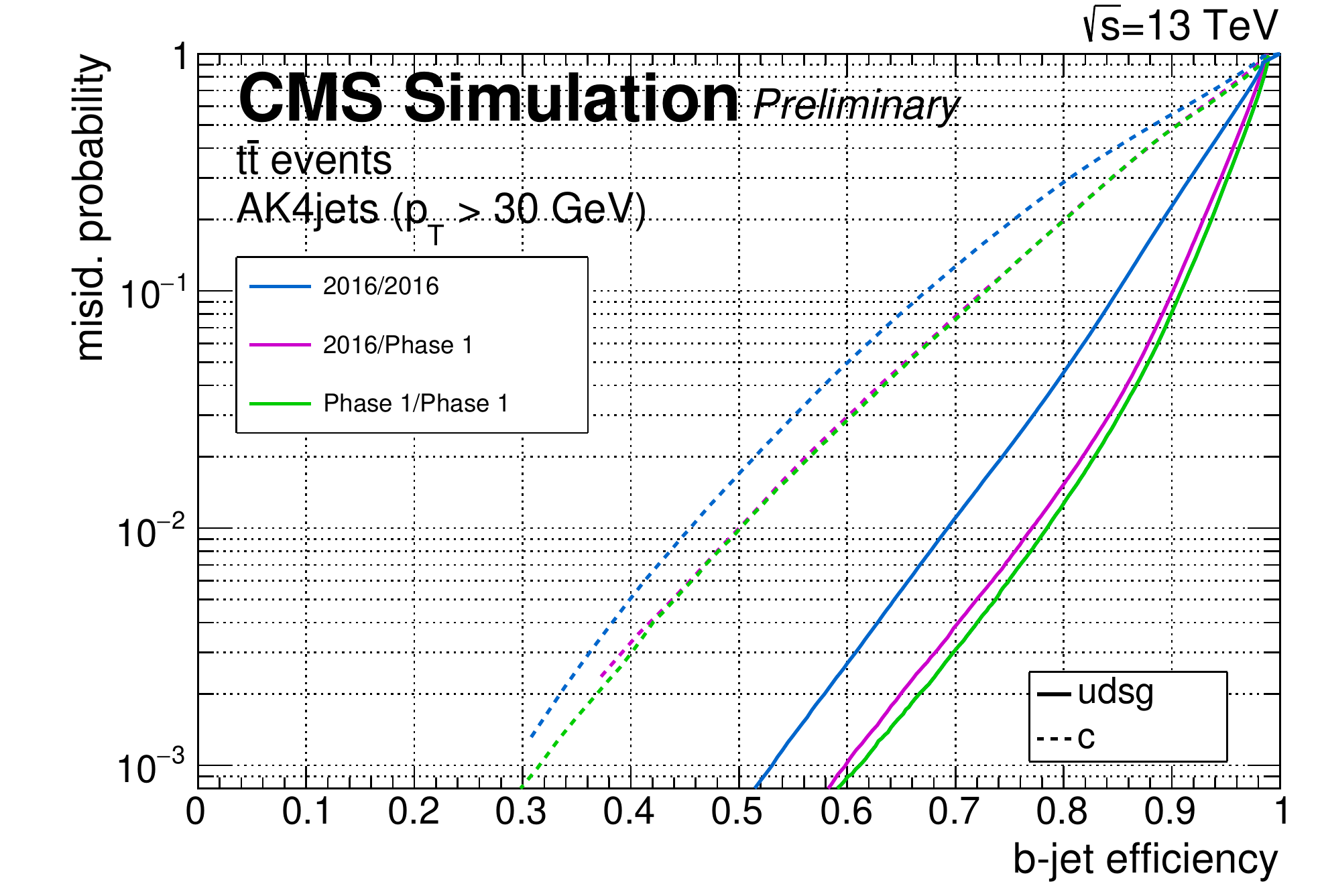}
\includegraphics[height=2in]{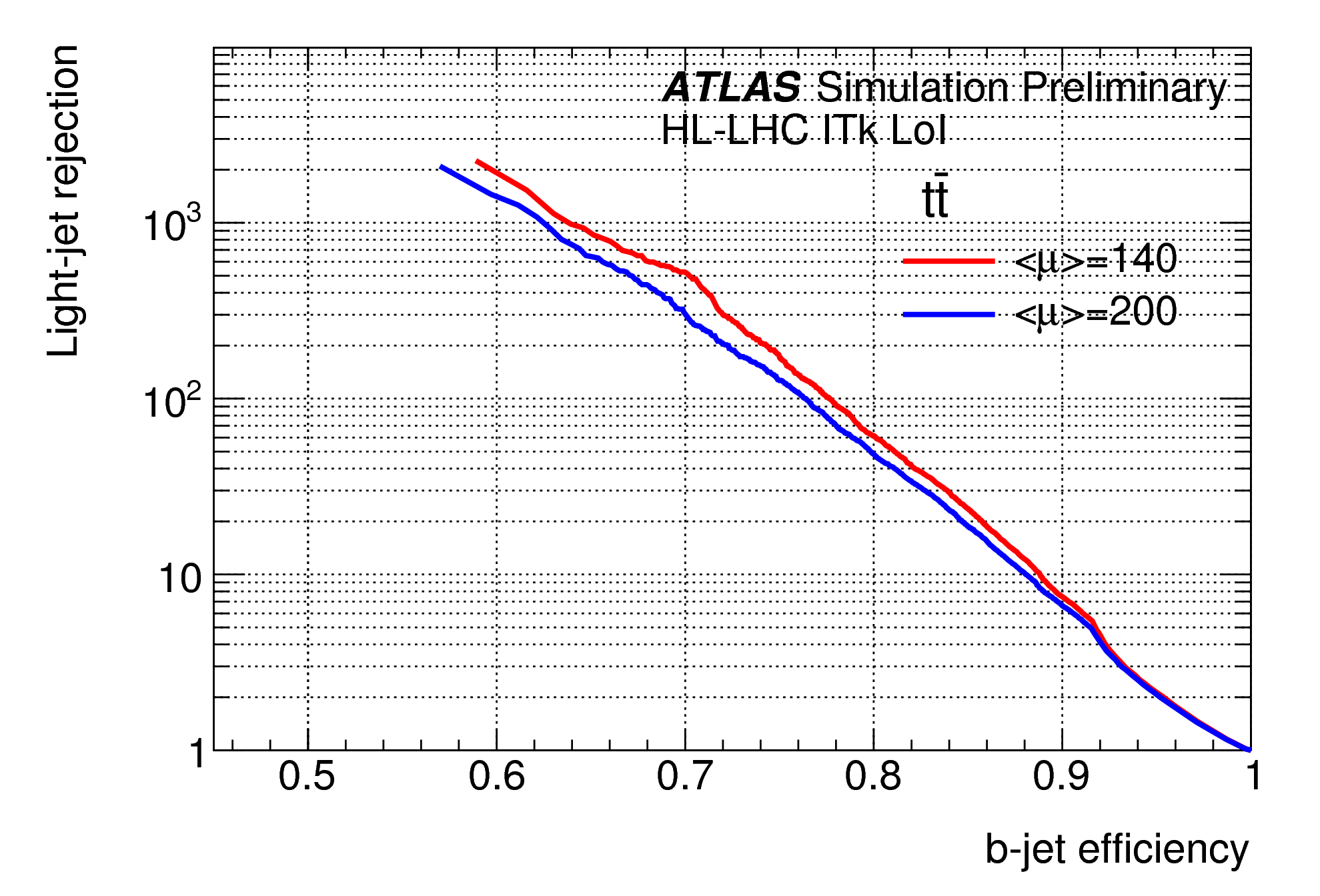}
\caption{Left: performance of the DeepCSV b tagging algorithm of CMS with the detector used in 2016 (blue line), with the new pixel detector and the 2016 training of the algorithm (magenta), and with the new pixel detector and a new dedicated training (green) \cite{CMS-ALG3}. Right: expected performance of b jet identification in ATLAS for two scenarios for the High Luminosity LHC phase \cite{ATLAS-UPG1}. }
\label{fig:upgrades}
\end{figure}

\section{Preparation for the Future Data Taking Periods}\label{sec:upgrades}

The ever increasing energies and instantaneous luminosities reached by LHC pose serious challenges to the reconstruction of the collected events. In order to cope with the conditions registered during Run2, the CMS experiment has installed during winter 2017 a new pixel detector \cite{CMS-UPG1} whose additional layer closer to the beam spot is expected to provide a better resolution on the measured impact parameter of the reconstructed tracks. This in turn allows a substantial improvement in the performance of b jet identification, as shown in Figure \ref{fig:upgrades}, where the performance of the DeepCSV algorithm with the detector used in 2016 is compared with the one reachable with the new detector \cite{CMS-ALG3}. 

Major upgrades of the ATLAS and CMS detectors are planned to operate during the High Luminosity (HL) LHC phase. In particular, track detectors will be replaced with new devices with higher granularity, radiation robustness and extended coverage. First studies \cite{ATLAS-UPG1,CMS-UPG2} show that the b tagging algorithms can operate in the complex high pile-up environment expected during HL-LHC (see Figure \ref{fig:upgrades}). 

\section{Conclusions}

The identification of jets coming from the hadronization of b quarks is a fundamental tool in most physics analyses. Both the ATLAS and CMS Collaboration reached a significant improvement on their algorithms in Run2, and new promising ideas for further developments are being explored. Not only algorithms, but also the measurements of their performance on data had benefited from new ideas and of increased sample statistics in 2016. Techniques are being extended to cover more specific topologies becoming ever more important with the increase of the LHC collisions center-of-mass energy. More challenges are ahead in view of the future data taking periods: the experiments are already working to maintain b tagging a successful tool in the next decades.


\end{document}